\documentclass[pra,twocolumn,amssymb]{revtex4}

\usepackage{amsmath}
\usepackage{amssymb}
\usepackage{bbm}
\usepackage{mathtools}
\usepackage[normalem]{ulem}
\usepackage{dsfont}
\usepackage{mathrsfs}
\usepackage{cancel}

\usepackage{braket}

\usepackage[utf8]{inputenc}
\usepackage[T1]{fontenc}

\def\la{\langle}
\def\ra{\rangle}

\def\bg{\begin{equation}\begin{gathered}}
\def\eg{\end{gathered}\end{equation}}

\def\B#1{\!\left(#1\right)}
\def\BB#1{\!\left[#1\right]}

\def\be{\begin{equation}}
\def\ee{\end{equation}}

\def\bee{\begin{equation*}}
\def\eee{\end{equation*}}

\def\ba{\begin{equation}\begin{aligned}}
\def\ea{\end{aligned}\end{equation}}

\def\Var{{\rm Var}}

\def\sing{\text{sing}}

\makeatletter
\newcommand{\pushright}[1]{\ifmeasuring@#1\else\omit\hfill$\displaystyle#1$\fi\ignorespaces}
\newcommand{\pushleft}[1]{\ifmeasuring@#1\else\omit$\displaystyle#1$\hfill\fi\ignorespaces}
\makeatother

\usepackage[ddmmyyyy]{datetime}
\makeatletter
\def\Dated@name{}
\makeatother

\begin{document}

\title{
Evidence from on-site atom number fluctuations for a quantum 
Berezinskii-Kosterlitz-Thouless  transition in the one-dimensional Bose-Hubbard model 
}
\author{Mateusz Łącki and Bogdan Damski}
\affiliation{Institute of Theoretical Physics,
Jagiellonian University, {\L}ojasiewicza 11, 30-348 Krak\'ow, Poland}
\begin{abstract}
We study the one-dimensional Bose-Hubbard model describing the superfluid-Mott insulator quantum phase transition 
of cold atoms in  optical lattices. 
We  show that derivatives of the variance of the on-site atom number occupation,  
computed with respect to the parameter driving the transition,
have  extrema that are located off the critical point even in the thermodynamic limit.
We discuss whether  such extrema
provide solid  evidence of
the quantum Berezinskii-Kosterlitz-Thouless transition taking place in this system.
The calculations are done for systems with 
the mean number of atoms per
lattice site  equal to either one or two.
They also  characterize the nearest-neighbor correlation 
function, which is typically discussed in the context of 
time-of-flight images of cold
atoms.

\end{abstract}
\date{\today}
\maketitle

\section{Introduction}
\label{Introduction_sec}

Over the last two decades, 
we saw an explosion of activities in 
theoretical and experimental studies of 
cold  atoms in external  (oftentimes periodic)  potentials 
\cite{LewensteinAdv,BlochRMP2008,Qsim1,KubaReview,KrutitskyPhysRep2015,BlochSci2017}. 
Decisive  motivation for these efforts  
came  from the observation  that such systems  
may  provide
unique insights into outstanding 
problems of  condensed matter physics. 
Such a presumption  follows from well-known facts that (i) 
various lattice geometries can be optically imposed on cold atoms
(one-, two-, and three-dimensional, square, triangular, etc.); 
(ii) different types of interactions can  be encountered  in such systems
(on-site, nearest-neighbor, long-range, etc.); (iii)  parameters 
characterizing them can be typically  
tuned over a vast range of values, which 
should allow for reaching the strongly-correlated quantum regime.

As a result,  tens of different condensed matter models,
which can  be neither  analytically solved nor efficiently numerically  
simulated,
were  conjectured  to
be experimentally accessible in cold atom systems. 
In the context of our work, 
those   undergoing  a quantum phase transition
are of special interest \cite{PiersNature2005,SachdevToday,SachdevBook,ContinentinoBook}. Among them
various Bose-Hubbard-like models
can be  most naturally approached  with cold atoms,
which is comprehensively discussed in reviews
\cite{LewensteinAdv,KubaReview,KrutitskyPhysRep2015}.

Suppose now that a strongly-correlated state 
of those atoms is created. The following question then 
arises: What experimentally-accessible observables
can be used for getting insights  into its properties?

To proceed with the  discussion of this question, it
should be said  that the most ubiquitous approach to experimental 
probing  of the state of 
cold atoms 
is based on the time-of-flight imaging technique, where
one turns off external fields keeping atoms in place. 
Atoms   fly away from each
other and then  their spatial
distribution is recorded, which is reviewed in  Ref. \cite{BlochRMP2008}. 
Similar  insights can 
be also  obtained through quantum gas microscope techniques,
where one probes  in-situ   distribution of atoms 
in individual lattice sites (see 
Refs. \cite{OttReview2016,TakahashiReview2020} for  reviews).

The former approach allows  for determination of two-point
correlation functions, out of which the nearest-neighbor one, 
i.e. the expectation value of the tunneling operator, is of special 
interest and will be commented upon below (see e.g. Ref. \cite{3DBHexp} for 
relevant recent experimental work). 
The latter approach  gives direct  insights into 
local atom number fluctuations, out of which  
the variance of the on-site atom number occupation can be determined.
Alternatively, one may employ the   atom-number-projection
spectroscopy for
measuring the variance, which is also discussed in above-mentioned 
Ref.  \cite{3DBHexp}.
Having said all that, 
it is now  natural to ask what imprints  of a quantum phase transition
can be seen in these  observables?

We have addressed such a question in systems 
described by two- and
three-dimensional Bose-Hubbard models.
Namely, it was shown in Refs. \cite{BDSciRep2016,BDSciRep2019}   that derivatives
of   both the variance and  the nearest-neighbor
correlation function, 
computed with respect to the parameter driving the transition,
have extrema, which 
can be used for  localization of critical points of such models.

The questions we are now interested in are  the  following. 
Can we gain  unique insights, via above-mentioned observables, 
into the very nature of the quantum phase
transition of the  one-dimensional (1D)
Bose-Hubbard (BH) model? 
How the results for this model 
differ from the ones obtained in its higher dimensional counterparts?

The outline of this paper is the following. 
The model that we study is presented in Sec. \ref{Model_sec}. 
Numerical simulations, for systems with the mean number of atoms per lattice
site equal to one, are 
discussed  in Secs. \ref{Numerical_sec} and \ref{BKT_sec}. 
The summary of our work is provided in Sec. \ref{Summary_sec}.
There are also two appendices. Appendix \ref{Double_sec} extends our studies 
from Secs. \ref{Numerical_sec} and \ref{BKT_sec}
to systems with two atoms per lattice site, 
whereas
Appendix \ref{Numerics_sec} presents technical details of our numerical
simulations.

\section{Model}
\label{Model_sec}
We study ground states of the 1D BH model with open boundary conditions.
Its Hamiltonian, expressed in the  unit of the
on-site interaction energy, is given by
\be
\begin{aligned}
&\hat{H}= -J \sum_{i=1}^{M-1}
\B{\hat{a}_{i+1}^\dag \hat{a}_i + 
\hat{a}_i^\dag\hat{a}_{i+1}
}
+\frac{1}{2} \sum_{i=1}^M\hat{n}_i\B{\hat{n}_i-1}, \\
&[\hat{a}_i,\hat{a}_j^\dag]=\delta_{ij}, \ [\hat{a}_i,\hat{a}_j]=0, 
\  \hat n_i=\hat{a}_i^\dag\hat{a}_i,
\end{aligned}
\label{H}
\ee
where 
$\hat a_i^\dag$ ($\hat a_i$) creates 
(annihilates) an atom in the $i$-th lattice site, 
$J$ is the nearest-neighbor tunneling coupling,
and $M$ is the number of lattice sites ($M\to\infty$ 
is assumed  in this section).
Physical realization of such a model, envisioned in 
seminal work \cite{JakschPRL1998}, 
asks for placement of cold atoms in an optical box trap
superposed onto an optical lattice. This 
should be possible due  to  recent unprecedented 
experimental  advances
in studies of  box-trapped gases,
which are summarized
in  latest review \cite{Hadzibabic2021}.

Assuming that the  lattice is 
filled with $N$ atoms, one defines 
the filling factor 
\be
n=N/M
\ee
being of key importance during discussion of many-body phases of
the 1D BH model. Namely,  at fixed integer $n$  
such a model undergoes the superfluid--Mott
insulator quantum phase transition 
\cite{FisherPRB1989,KrutitskyPhysRep2015}. 
Such a transition  lies in the 
universality class of the  two-dimensional classical
XY model \cite{remark}. 
This means that it is  
the quantum Berezinskii-Kosterlitz-Thouless (BKT) transition, which in the
classical context 
was  described in seminal  works \cite{BerezinskiI,BerezinskiII,KT}
(see Ref. \cite{KTreview} for a recent review).

The system 
is in the Mott insulator (superfluid)  phase when  $0\le J<J_c$ ($J>J_c$).
For the unit filling factor, being of interest in the 
main body of this paper, the critical point is
located at   $J_c\approx0.3$, which is
more than three times  larger than 
the mean-field prediction \cite{StoofPRA2001}.
A thorough summary of theoretical efforts leading to
such a value is presented in  Ref. \cite{KrutitskyPhysRep2015}.

The question now is how the BKT character of the
superfluid--Mott
insulator phase transition of (\ref{H}) can be probed in cold atom experiments.
We suggest that this can be done by taking a closer look 
at 
either  the  variance of
the on-site atom number occupation
\be
\Var(J)=\langle J|\hat n_i^2|J\rangle - n^2
\label{var}
\ee
or  the nearest-neighbor
correlation function
\be
C(J)=\la J| 
\hat a^\dag_{i+1}\hat a_i 
+
\hat a^\dag_i\hat a_{i+1} 
|J\ra,
\ee
where $|J\ra$ denotes  the ground state of (\ref{H}).
 Equivalence of the physical content of $\Var(J)$ and $C(J)$, 
 whose derivatives with respect to the parameter
driving the transition will be extensively discussed 
below, comes from the mapping
\be
\frac{d}{d J}\Var(J)=2J\frac{d}{d J} C(J),
\ee
which can be easily found via the Feynman-Hellmann theorem.

A more insightful result coming from such a theorem is that 
\be
\frac{d}{d J}\Var(J)=-2J\frac{d^2}{d J^2}{\cal E}(J),
\label{dVar}
\ee
where ${\cal E}(J)$ is the ground-state energy per lattice site.
This simple identity provides a link between derivatives of the variance 
and physics of BKT  transitions.

It is so because the singular part of 
the ground state energy density
is expected to be well-approximated 
by the  BKT-type expression on 
the Mott insulator side of the transition 
\cite{FisherPRB1989}
\be
{\cal E}_\sing(J)\approx
A\exp\B{-\frac{2B}{\sqrt{J_c-J}}}\sim\xi^{-2}(J),
\label{Esing}
\ee
where $\xi$ is the 
correlation length while 
$A$ and $B$ are  some non-universal constants. 
We 
focus our attention on the Mott insulator phase 
in this work.

Finally, to place our studies in a larger 
setting, we have
the following comments.

First,  in the context of classical  phase transitions, where 
the BKT theory is typically discussed \cite{KTreview}, (\ref{Esing})  
describes the singular part of the free energy density (see e.g.
Refs. \cite{FinotelloPRB1989,FinotelloPRL1993}). 
  Having said that, we see from (\ref{dVar}) that $d\Var/dJ$ is the 
  exact quantum
analog of the specific heat, whose behavior 
is 
of special interest in the  classical context. 
This remark follows from the fact that 
the specific heat per lattice site can be 
written as $-Td^2{\cal F}/dT^2$,
where $T$ is temperature and 
$\cal F$ is the free energy per lattice site \cite{Baxter}. 

\begin{figure}[t]
\includegraphics[width=\columnwidth,clip=true]{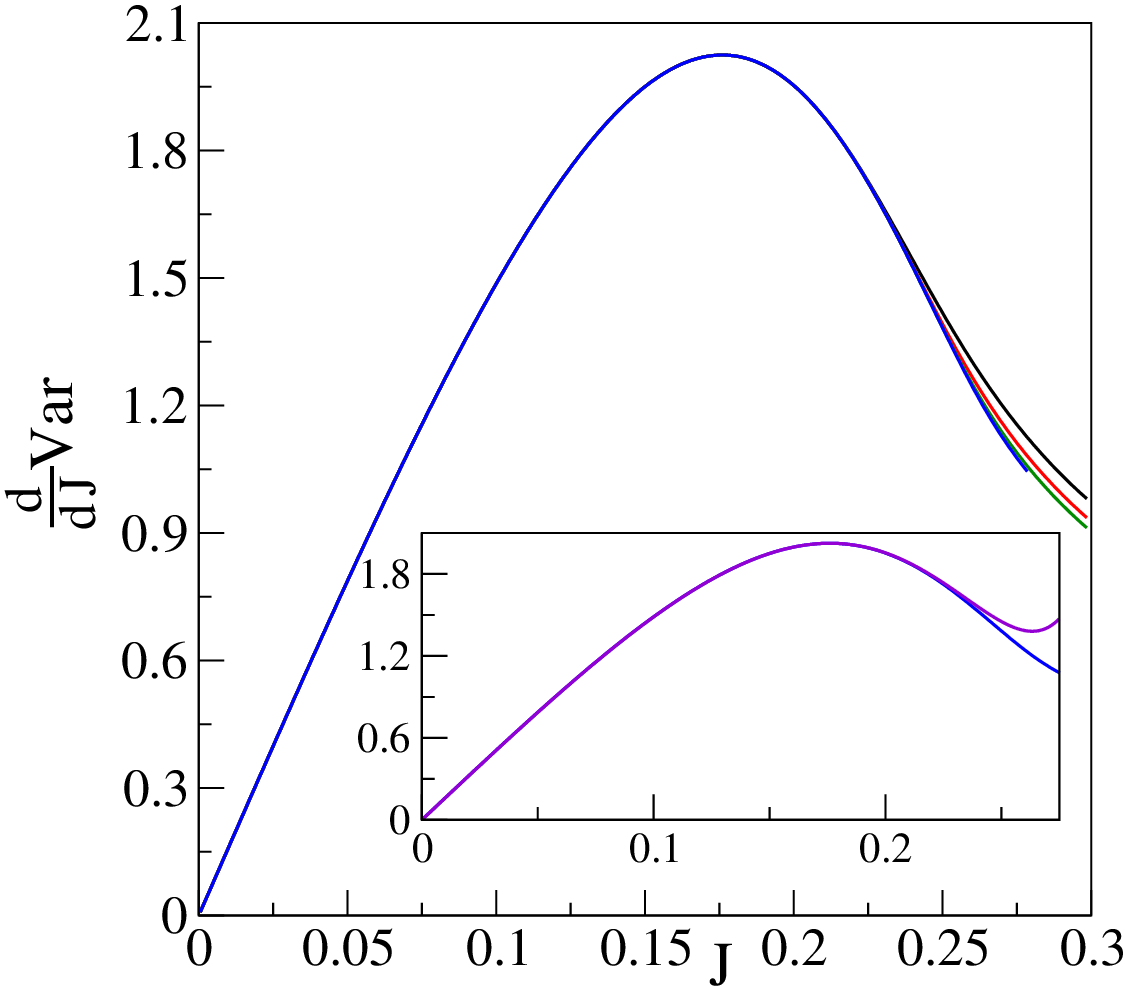}
\caption{The first derivative of the variance 
of the on-site atom number occupation 
for the unit  filling factor ($n=1$).
Main plot: numerical results 
for system sizes $M=100$ (black), $200$ (red), $400$ (green), and $800$
(blue). 
Inset: the violet curve depicts the perturbative result
obtained from (\ref{var_n=1}) while 
the blue one  shows     $M=800$ 
data from the main plot. The two curves are practically indistinguishable 
for $J$ smaller than about $0.2$.
}
\label{d1n1}
\end{figure}

Second, insights into 
BKT physics of the 1D BH model
can be also obtained from 
the single-particle energy gap, 
 which is proportional to $\xi^{-1}(J)$.
In addition to that, one may 
also study 
two-point correlation
functions
\be
\la J| 
a^\dag_{i+r}\hat a_i 
+ 
\hat a^\dag_i\hat a_{i+r} 
|J\ra,
\label{Cr}
\ee
which for $r\gg1$
are  expected 
to exhibit 
the algebraic $r^{-1/4}$ 
decay at the BKT critical point 
(we overlook an 
essentially 
unobservable logarithmic correction to such a decay law, see e.g. 
Ref. \cite{ChaikinBook}).
Numerical studies of the former (latter) quantity 
can be found in 
Ref. \cite{RigolPRA2013} 
(Refs. \cite{MonienPRB1998,MonienPRB2000,KubaDom}; see also Ref.
\cite{LeHurPRL2012}).
As far as experiments are concerned, 
 it is  unclear to us  whether 
one can  measure these quantities accurately-enough
for getting conclusive insights into BKT physics.
We mention in passing that more ``exotic'' 
physical quantities, 
providing  insights  complementary
to the ones 
delivered by $d\Var/dJ$,
will be commented upon  in Secs. \ref{BKT_sec} and \ref{Summary_sec}.

Third, 
experimental studies of  BKT physics  in cold atom setups
were initiated by seminal work  \cite{HadzibabicNature2006}. 
To the best of our knowledge, however, 
they were   restricted to 
two-dimensional cold  gases, where the classical BKT
transition takes place 
(see e.g. Refs. \cite{HadzibabicNature2006,ChinNature2011,Homog1,HadzibabicNature2021}
 reporting experiments with either harmonically trapped or 
 homogeneous  Bose gases; see 
 Ref. \cite{HadzibabicDalibardReview} for a review).
Note  that 
the quantum character of the BKT transition,
  1D geometry, and the  periodic 
 lattice potential  in our system  strikingly 
 contrast with the 
 properties of the systems explored in these experimental works.

\section{Numerical simulations}
\label{Numerical_sec}

Model  (\ref{H}), just as its two- and three-dimensional 
incarnations, is not exactly solvable. As a result, its theoretical studies
are oftentimes carried out  via numerical
simulations. 
Our numerical simulations are presented in Figs. \ref{d1n1}--\ref{d3n1},
showing the first, second, and third derivative of the variance
in the Mott insulator phase (see 
Appendix \ref{Numerics_sec} for technical details).

The most striking features seen on these 
figures are  the extrema, whose location 
\be
\begin{array}{c|c|c}
  & \text{minimum} & \text{maximum}   \\
  \hline
  d\Var/dJ 	& 	& 	0.176 \\ 
  \hline
  d^2\Var/dJ^2  &  0.247 & 	  \\  
  \hline
  d^3\Var/dJ^3  & 0.198   & 0.266  
\end{array}
\label{tab1}
\ee
is listed here  for the largest system that we
have numerically studied ($M=800$). 

The broad  maximum of $d\Var/dJ$,
 depicted in Fig. \ref{d1n1}, is 
the quantum equivalent of the so-called 
non-universal specific 
heat peak that was predicted by the BKT theory 
(see
e.g. Ref. \cite{ChaikinBook}, where such 
terminology
is used in the classical context). 
  Its off critical point location 
  nicely illustrates the peculiar nature of  BKT 
  transitions. Indeed, in 
  two- and three-dimensional 
  BH models, where  non-BKT
   transitions take place, 
  $d\Var/dJ$ has  
  maxima that  are  located 
  at the critical points  \cite{BDSciRep2016,BDSciRep2019}.

\begin{figure}[t]
\includegraphics[width=\columnwidth,clip=true]{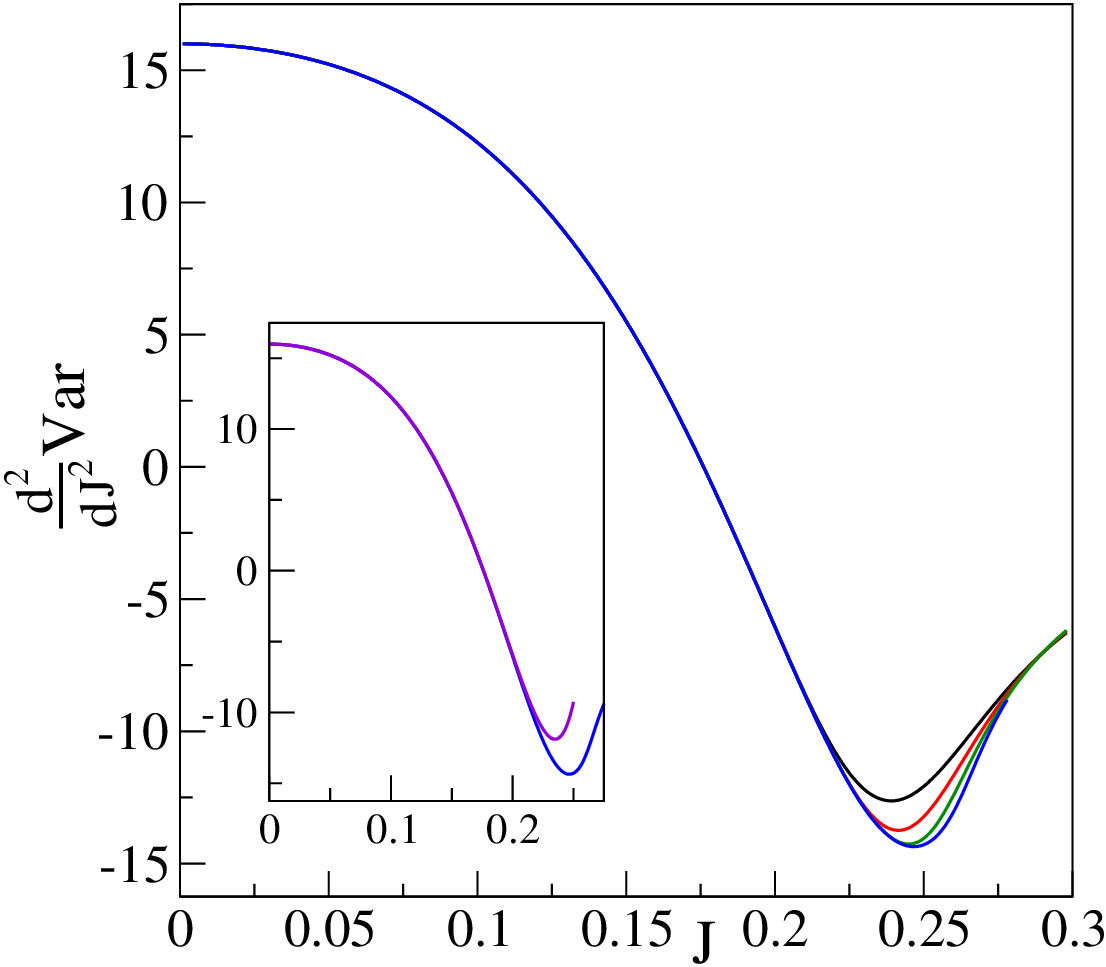}
\caption{The same as in Fig. \ref{d1n1} except we deal here with 
the second  derivative of the variance.
}
\label{d2n1}
\end{figure}

It is also evident from these figures that the non-universal 
contribution to 
the plotted quantities is by no means negligible at the critical point.
This conclusion follows from the observation that 
all derivatives of  (\ref{Esing}) vanish at the critical point 
(${\cal E}_\text{sing}$ is essentially 
singular at $J_c$). 
Thus,  
 derivatives of the variance  at the critical point are
entirely determined by the  non-universal 
component of the ground state energy density. They are 
clearly far from being negligible 
there, which is seen  after extrapolation 
of the data from Figs. \ref{d1n1}--\ref{d3n1} to 
$J=J_c\approx0.3$ \cite{remarkNext1}.
Similar situation is found in classical BKT transitions, 
where the
specific heat near 
critical points is 
known to be dominated by 
non-universal contributions \cite{KTreview}.

We also note  that finite-size effects  are most evident 
near the maximum of the third derivative (Fig. \ref{d3n1}).
This  is related to  the fact that the 
correlation length of the infinite 1D BH model
at $J=0.266$  is equal to about  four hundred \cite{MarekPRX2018},
which is only a factor of two smaller than the 
largest system size that we have numerically studied.

Then, we compare numerical simulations to 
analytical results  following from 
\be
\begin{aligned}
\Var(J)&= 8 J^{2} -24 J^{4} -\frac{2720}{9} J^{6} +\frac{70952}{81} J^{8}
-\frac{176684}{81} J^{10}\\& +\frac{431428448}{6561} J^{12}
+\frac{104271727762891}{330674400} J^{14}
\\&+\frac{32507578587517774813}{3888730944000} J^{16} + O\B{J^{18}},
\end{aligned}
\label{var_n=1}
\ee
which was obtained via the Rayleigh-Schr\"odinger perturbative expansion in the 
tunneling coupling for an infinite system subjected to the unit filling factor
constraint \cite{BDNJP2015}. We mention in passing that Ref. \cite{BDNJP2015}
comprehensively presents high-order  perturbative
 studies of the Mott insulator phase of the  1D 
BH model  
(see also Refs. \cite{MonienPRB1996,MonienPRB1999,MonienArxiv,EckardtPRB2009,TrivediPRA2009,KnapPRA2012}
for related albeit lower-order investigations).

\begin{figure}[t]
\includegraphics[width=\columnwidth,clip=true]{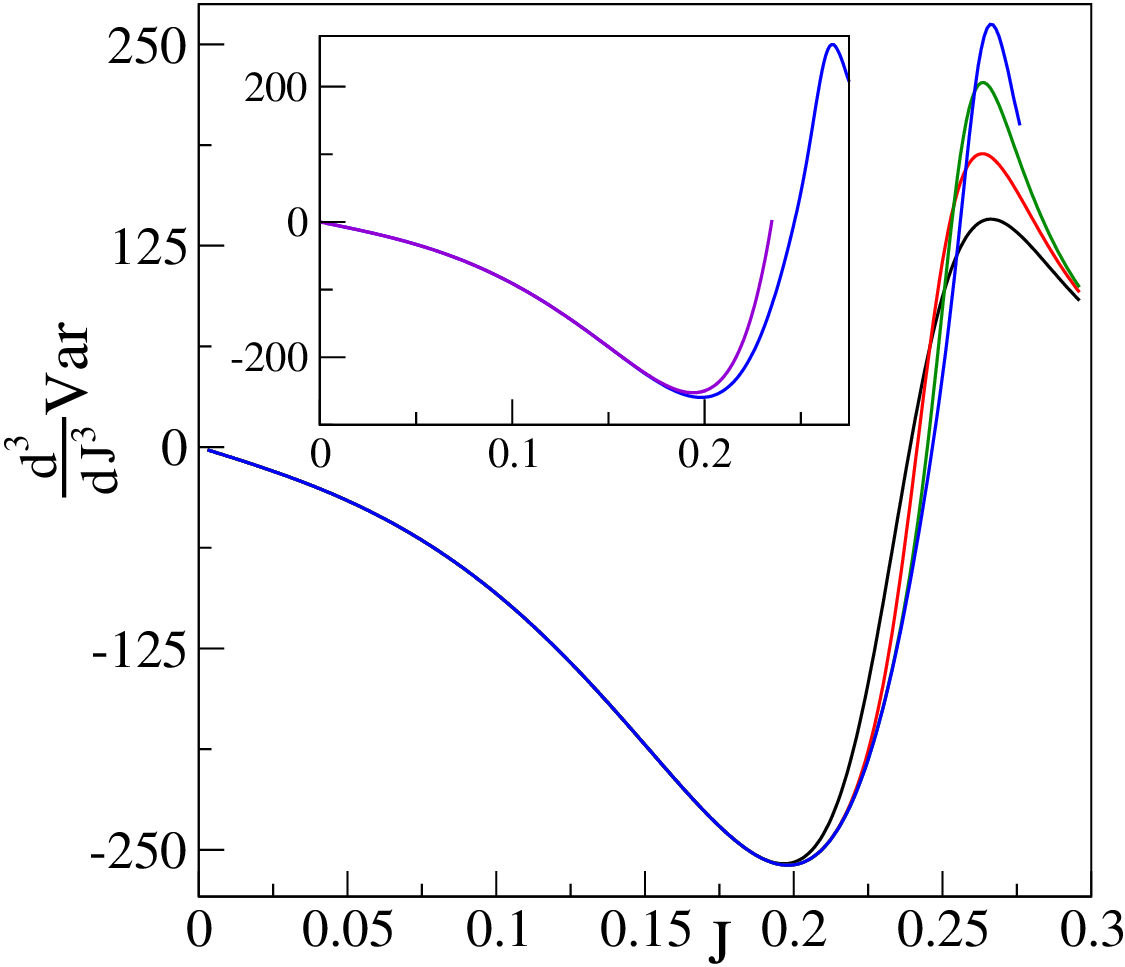}
\caption{
The same as in Fig. \ref{d1n1} except we deal here with
the third   derivative of the variance.
}
\label{d3n1}
\end{figure}

To begin, we take a look at  positions of  extrema
  following from  (\ref{var_n=1}). They  are  given by   
\be
\begin{array}{c|c|c}
  & \text{minimum} & \text{maximum}   \\
  \hline
  d\Var/dJ      &       &       0.176 \\ 
  \hline
  d^2\Var/dJ^2  &  0.235 &        \\  
  \hline
  d^3\Var/dJ^3  & 0.194   & 
\end{array}\,,
\ee
which quite accurately reproduces
all but one result reported in   (\ref{tab1}).
The position of the maximum of the third derivative is missing
here because such a maximum is absent in
 the third derivative of (\ref{var_n=1}).

Next, we note that a very good agreement 
between numerics  and perturbative results is seen 
for $J$ less than about 
$0.2$. This is sufficient for excellent (good)
analytical characterization of the maximum  (minimum) in Fig. \ref{d1n1} (Fig.
\ref{d3n1}). However, despite the high order of expansion (\ref{var_n=1}),
the shape of the minimum in Fig. \ref{d2n1} is only reasonably reproduced
by the perturbative formula
while  the maximum in Fig. \ref{d3n1} is not captured by it, which we have
already mentioned.
This is presumably so because these two features are located at
so large $J$ that  a  higher-order expansion is needed.
For example, already near $J=0.2$, we can infer 
from the  numerical data that the  
low order of the expansion, 
rather than finite-size effects, is responsible for 
discrepancies between 
 numerics and analytics (the larger the system size is, the 
bigger they are).

The question now is how we can actually argue that the above-discussed 
numerics  provides evidence of the quantum BKT transition
taking place in our system. This  brings us to the next section.

\section{BKT fit}
\label{BKT_sec}

The idea here is to fit 
\be
\frac{d}{dJ}\Var(J)= -2J\frac{d^2}{dJ^2}\BB{A\exp\B{-\frac{2B}{\sqrt{J_c-J}}}}
+C J + D J^2
\label{dVar_ABCD}
\ee
to numerics from Fig. \ref{d1n1},  use so determined expression 
to compute higher derivatives of the variance, and finally 
to compare such obtained results for 
$d^2\Var/dJ^2$ and $d^3\Var/dJ^3$
to  numerics presented in Figs. \ref{d2n1} and \ref{d3n1}, 
respectively.
Two  remarks are in order now.

First, we set $J_c=0.3$ in (\ref{dVar_ABCD}),
taking such a value from Ref. \cite{KrutitskyPhysRep2015}.
The fitting procedure yields the $A$, $B$, $C$, and $D$
coefficients. 
It will be applied to all data points
that we have, which represent
$d\Var/dJ$ in the Mott insulator phase.
Such a choice of the 
range of  $J$'s is motivated by the 
fact that the features that  we try to 
capture, such as the maximum from Fig. \ref{d1n1}, 
are not necessarily located near the critical point. 
Moreover, 
we reduce a bit arbitrariness of the fitting procedure
by avoiding  fine-tuning of the domain of (\ref{dVar_ABCD}).

Second, the exponential term in (\ref{dVar_ABCD})
 comes from the universal BKT formula,
see (\ref{dVar}) and (\ref{Esing}).
The polynomial terms in (\ref{dVar_ABCD})
represent  
the non-universal contribution to $d\Var/dJ$ 
in the simplest  possible way.
This can be argued as follows.
The constant, $J$-independent
term is skipped   as we expect 
from perturbative expansions that $d\Var/dJ$ vanishes at $J=0$.
Both  linear and quadratic terms in $J$ are needed for capturing 
the overall parabolic shape of the data from Fig. \ref{d1n1}.
Omission of the cubic, quartic, etc. terms in $J$ 
reduces the number of free parameters to  minimum.
Similar fitting schemes were explored in the classical 
context in  Refs. \cite{FinotelloPRB1989,FinotelloPRL1993}.

\begin{figure}[t]
\includegraphics[width=\columnwidth,clip=true]{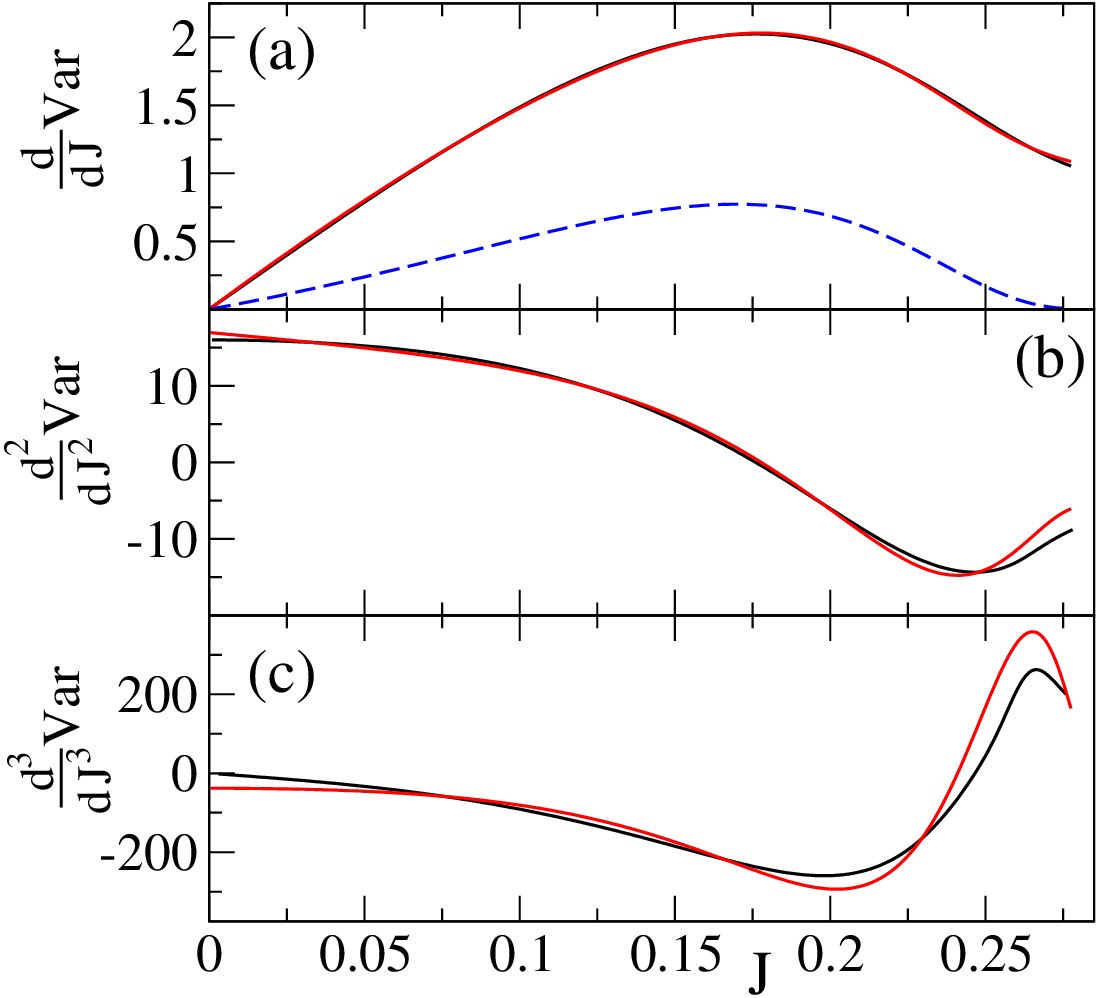}
\caption{
Comparison between 
numerics  and the BKT fit
discussed in Sec. \ref{BKT_sec}.
Black lines show numerics   for the
largest system that we consider ($M=800$). 
Red lines follow from 
 (\ref{dVar_ABCD}) 
evaluated with the coefficients from (\ref{ABCD}).
 The dashed blue line in panel (a) depicts the universal 
part of the fitted function, i.e., the exponential 
contribution from  (\ref{dVar_ABCD}).
All results are for the unit filling factor ($n=1$).
}
\label{kt_n=1}
\end{figure}

The fitting has been done
with  the NonlinearModelFit function
from Ref. \cite{Mathematica}. It  yielded 
\be
\begin{array}{c|c|c|c}
 A & B & C & D \\
 \hline
 -12.5(2) & 1.465(3) & 12.8(1) & -32.1(4)
\end{array}\,,
\label{ABCD}%
\ee
where one standard error is 
listed  in the brackets. 
All data from Fig. \ref{d1n1}, 
for the $M=800$ system, has 
been used for the fit.

Out of these four fitting results, only 
the $B$ parameter can be 
compared to the former 
studies. Namely, 
it 
was  extracted from 
 numerical data for  the single-particle energy gap, correlation length, ground
state fidelity, and fidelity susceptibility
of the 1D BH model \cite{RigolPRA2013,MarekPRX2018,MarekPRB2019}.
Those  studies  
estimated it   at $1.59(3)$,  $1.61(4)$, $1.72(1)$, and $1.84(5)$,
respectively. Our result adds one more value to the table,
which does not seem to be solving  the puzzle of what the value of 
$B$ really is. 
Given the fact that  there is 
$25\%$ relative difference between the largest and 
the smallest reported value of $B$, 
further studies seem to be needed for 
tight
 estimation of this parameter.

The quality of the fit reported in (\ref{ABCD})
is depicted in Fig.  \ref{kt_n=1}a, where 
its good agreement  with  numerics is easily seen.
We also  separately  plot there 
the universal contribution to the fitted expression.
It is  peaked near the maximum of $d\Var/dJ$, where
it is  of the same order of magnitude
as the non-universal part of (\ref{dVar_ABCD}).
Moreover, as Fig.  \ref{kt_n=1}a reveals,
nowhere in the Mott insulator phase the universal 
contribution dominates over the non-universal one.
This observation  illustrates the curious nature of the 
studied  transition, so much different from what one
finds in  standard, non-BKT, transitions.

Next, we combine  (\ref{dVar_ABCD}) and (\ref{ABCD}) 
to compute higher  derivatives
of the variance and compare them to numerics in Figs. \ref{kt_n=1}b and \ref{kt_n=1}c. 
The agreement is good but not as good
as in 
Fig. \ref{kt_n=1}a. This is somewhat expected given the 
fact that 
we 
account for the non-universal part of the result 
with just a linear function ($C+2DJ$) in 
Fig. \ref{kt_n=1}b and 
a constant term ($2D$) in Fig. \ref{kt_n=1}c.

Having said all that, we can address 
the question posted by the 
end of Sec. \ref{Numerical_sec}. 
Namely, we see 
agreement between 
curves plotted in Fig. \ref{kt_n=1}
 as  solid  evidence that there is a quantum BKT transition
in our system. This remark should be especially convincing
if one  looks at Fig. \ref{kt_n=1}c, where whole $J$-dependence 
comes solely from the universal BKT formula properly
reproducing the shape of numerical data.

\section{Summary}
\label{Summary_sec}

We have 
discussed  how BKT physics of the 
superfluid--Mott insulator quantum 
phase transition of the 
1D BH model 
can be extracted from either  the
variance of the on-site  atom number occupation
or the nearest-neighbor correlation function.
It may seem surprising at first glance  that 
a clear signature of the BKT transition can be 
obtained from them. 
We say so because these two 
physical quantities
seem to be    featureless 
in the Mott insulator phase, where
we do calculations 
(see e.g. Ref. \cite{BDNJP2015}). 
Interestingly enough, 
this remark may explain the fact that we are 
unaware of any works  discussing them 
from  the BKT perspective. 

A clear link to  BKT physics appears when 
one considers the first derivative of the variance 
 with respect to
the parameter driving the transition.
It turns out that such a quantity is the 
exact 
quantum analog of  the specific heat (Sec. \ref{Model_sec}).
 Thus, by studying it, we get direct insights into 
 the quantum BKT transition from the  same perspective 
 from which  classical BKT transitions are oftentimes 
 discussed. The same can be said about the first 
 derivative of the 
 nearest-neighbor correlation function
 because it  is proportional to the
 first derivative of the variance (Sec. \ref{Model_sec}).

As far as experiments are concerned, both 
the variance and 
the nearest-neighbor correlation function
 can be measured (Sec. \ref{Introduction_sec}). 
We expect that 
it should be also possible to extract 
their derivatives out of experimental data. 
In this context, we would like to 
mention Ref. \cite{GerbierPRL2005}, 
where the derivative of experimentally-measured 
visibility 
of the time-of-flight interference 
pattern was
used for estimation of
critical points of the three-dimensional
BH model.
This work  
demonstrates  feasibility of studies 
of derivatives 
of quantities measured in cold atom experiments.

It should be mentioned,  however,
 that 
accurate computation of  derivatives 
of  experimental data 
 would presumably require  smoothing 
of such  data first
(e.g. by fitting some function to it).
Once this would be done, calculation of  
derivatives should be easy. 
We have
been able to avoid such a procedure in this work 
thanks to the high quality of 
numerical 
data that was differentiated (Appendix \ref{Numerics_sec}). 
However, 
in our former studies, where Quantum Monte Carlo 
simulations were employed \cite{BDSciRep2016,BDSciRep2019}, 
we  used the Pad\'e approximant fitting approach.

After this qualitative overview,
we would like to make the following comments.

First, we have
studied  systems with the
average number of atoms 
per lattice site equal to either one
(Secs. \ref{Numerical_sec} and \ref{BKT_sec}) or two (Appendix
\ref{Double_sec}). The latter case has been moved to the appendix 
because numerical results are 
similarly analyzed 
for both filling factors.

Second, we have discussed  a  scheme 
for extraction of  BKT physics out of
the above-mentioned observables (Sec. \ref{BKT_sec}).
By using it, we get to
know how much the universal part contributes to 
the quantities that we study. For example, 
how much it contributes to the quantum analog of the
so-called {\it non-universal}
peak of the specific heat, which is
depicted in 
Fig. \ref{kt_n=1}a.
We have shown that the universal component contributes 
to the peak about as much as the 
non-universal one.
As a result of that, 
it seems to us that the peak  in our system 
 actually appears  to be neither universal nor
non-universal.

Third, it should be said   that 
such a peak, to the best 
of our knowledge,
was  never  experimentally observed 
in cold atom systems undergoing a  quantum
BKT transition.

Fourth, we are aware of just one earlier 
theoretical work on 
the 1D BH model, 
where some quantum analog  of the 
specific heat peak was discussed from the BKT
perspective \cite{MarekPRB2019}.
Its experimental exploration, 
however,  asks for  the measurement of 
either  ground state fidelity or fidelity susceptibility.
As far as we understand it, it is unclear how to measure the former,
whereas the latter can be extracted 
from the  measurements of the 
spectral function \cite{GuEPL2014}.
It seems to us that the observables that we discuss are far more 
experimentally approachable.

Finally, we note that there are different versions of the 1D 
BH model 
 presumably undergoing a 
 quantum BKT  transition  \cite{LewensteinAdv,KubaReview,KrutitskyPhysRep2015}.  
We expect  
our ideas for studies  of BKT 
physics 
to be also   applicable to such 
models. We  hope that this work
 will stimulate experimental 
exploration of  cold-atom-based quantum BKT transitions.

\begin{figure}[t]
\includegraphics[width=\columnwidth,clip=true]{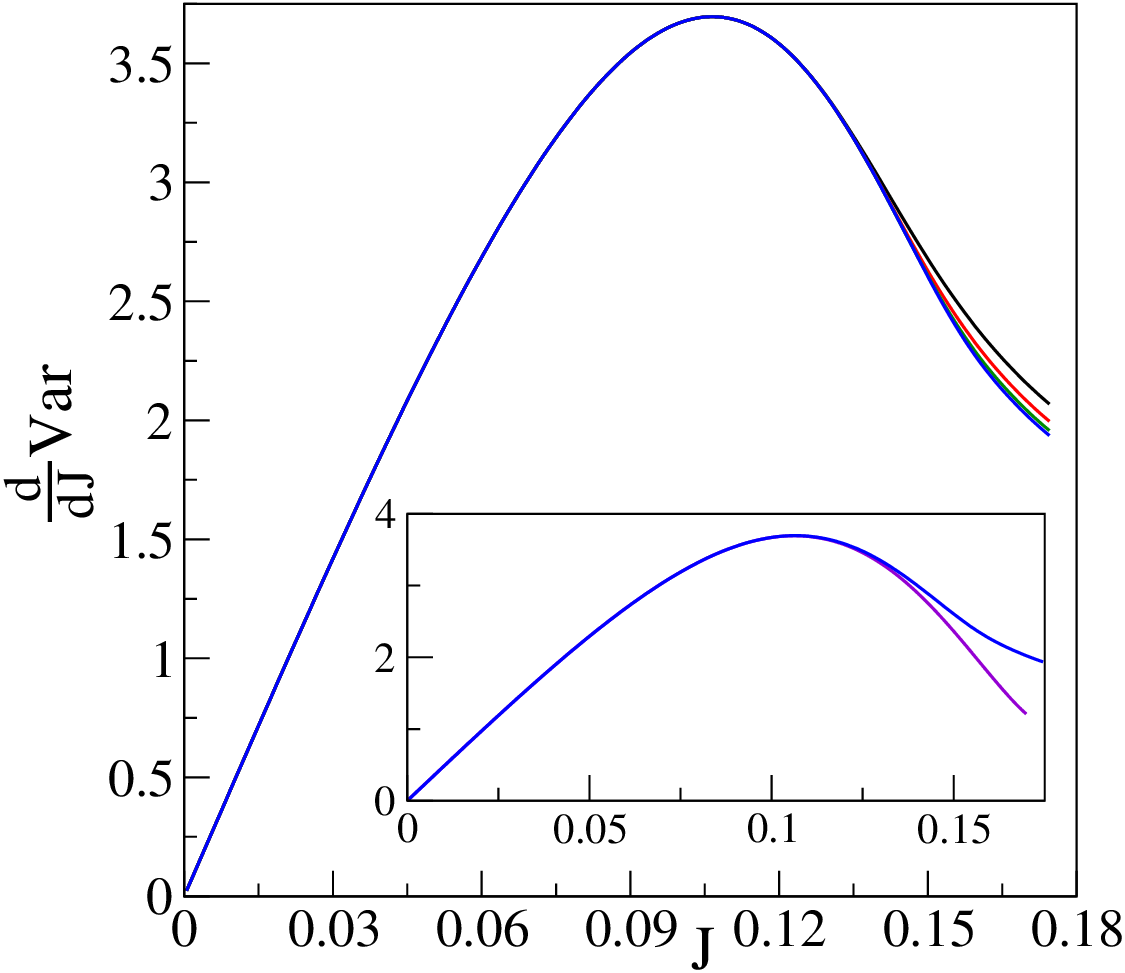}
\caption{
The first derivative of the variance
of the on-site atom number occupation
for the double   filling factor ($n=2$).
Main plot: numerical results
for system sizes $M=100$ (black), $200$ (red), $400$ (green), and $800$
(blue).
Inset: the violet curve depicts the perturbative result
obtained from (\ref{var_n=2}) while
the blue one  shows  the   $M=800$
data from the main plot. The two curves are 
nearly identical
for $J$ smaller than about $0.1$.
}
\label{d1n2}
\end{figure}

\section*{Acknowledgments}
We thank Marek Rams for both
useful discussions
and remarks about the manuscript. 
MŁ was supported by  the Polish National Science Centre (NCN) project 2016/23/D/ST2/00721. 
BD was supported by the Polish National Science Centre (NCN) grant 2016/23/B/ST3/01152.
Numerical computations in this work were supported in part by PL-Grid Infrastructure.

\appendix

\section{Double filling factor}
\label{Double_sec}
\begin{figure}[t]
\includegraphics[width=\columnwidth,clip=true]{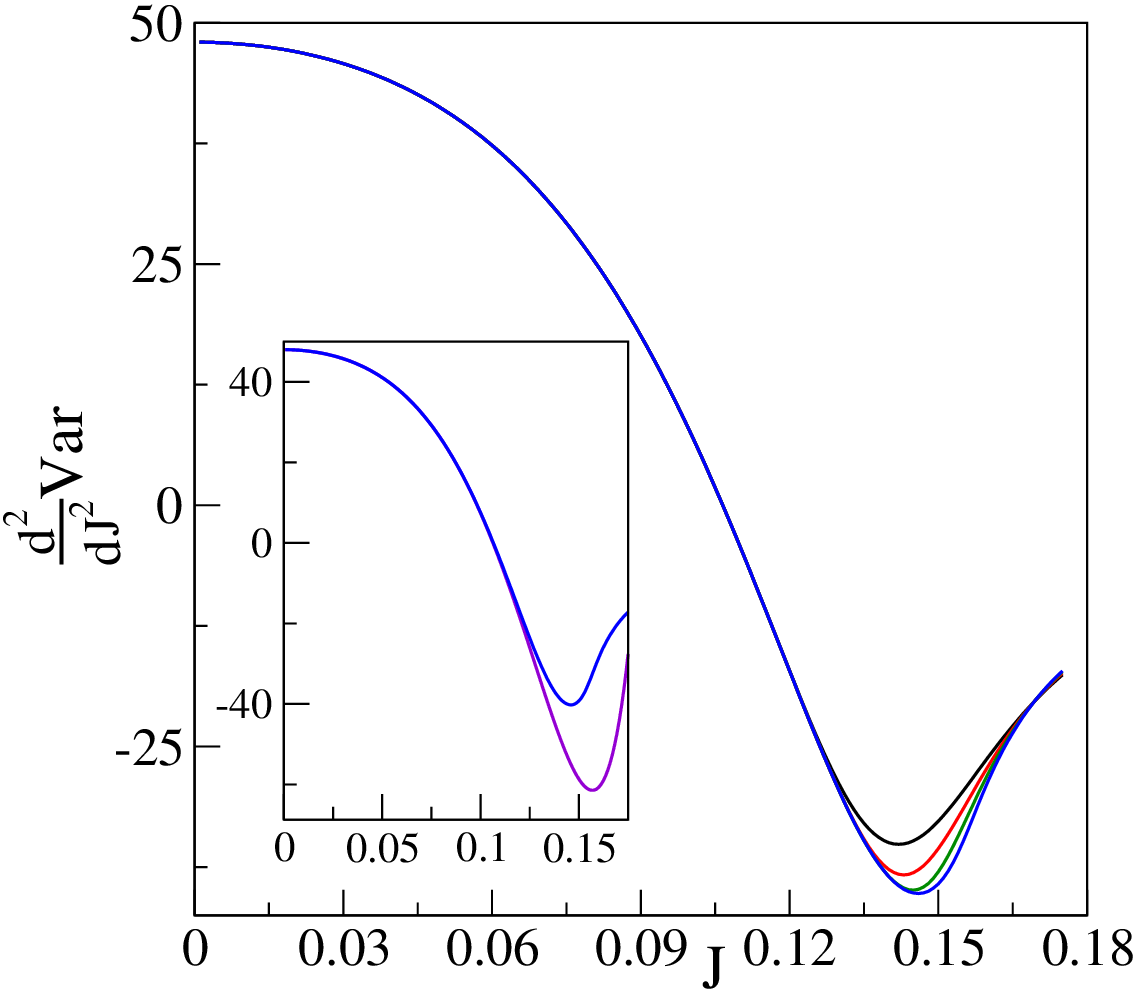}
\caption{
The same as in Fig. \ref{d1n2} except we deal here with
the second  derivative of the variance.
}
\label{d2n2}
\end{figure}

\begin{figure}[t]
\includegraphics[width=\columnwidth,clip=true]{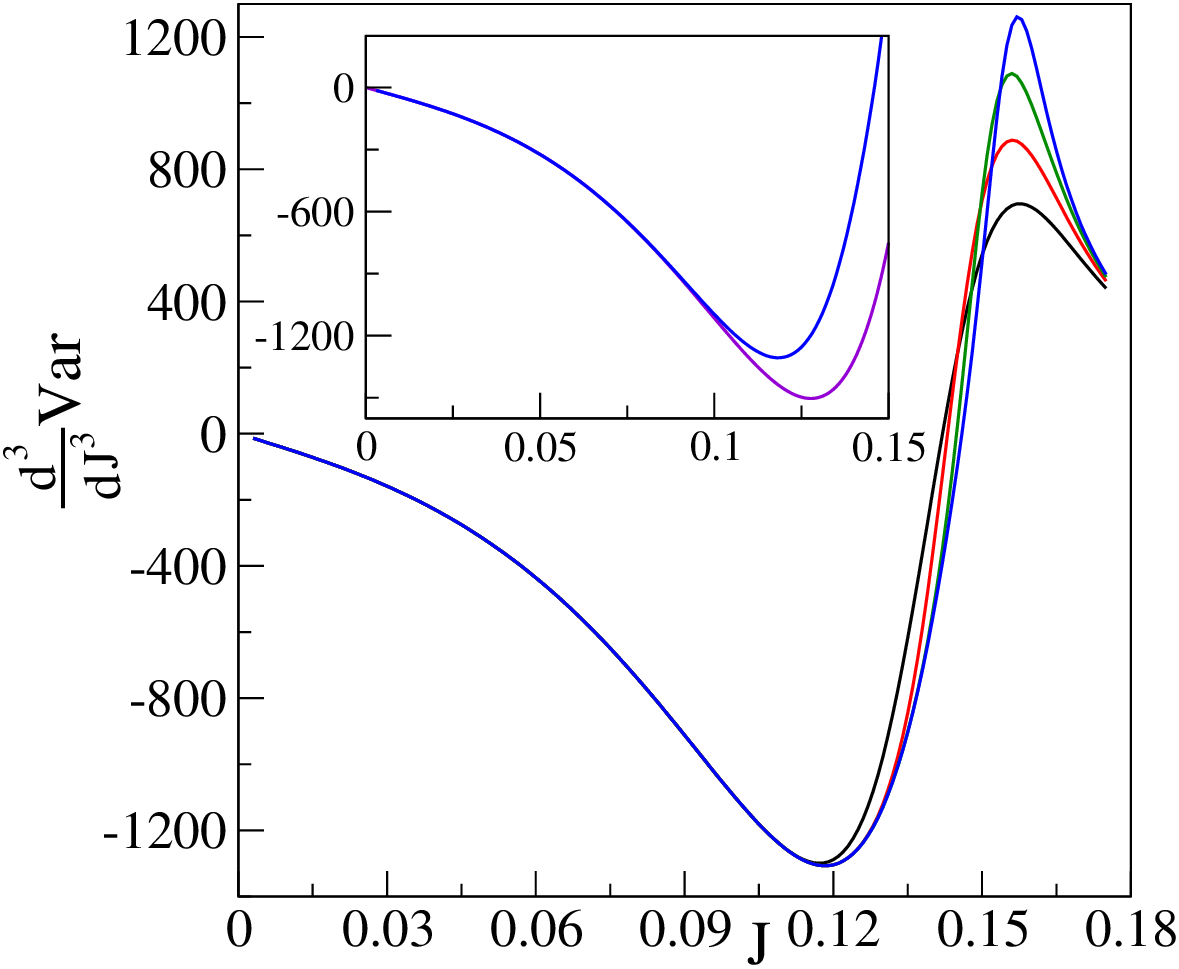}
\caption{
The same as in Fig. \ref{d1n2} except we deal here with
the third  derivative of the variance.
}
\label{d3n2}
\end{figure}

\begin{figure}[t]
\includegraphics[width=\columnwidth,clip=true]{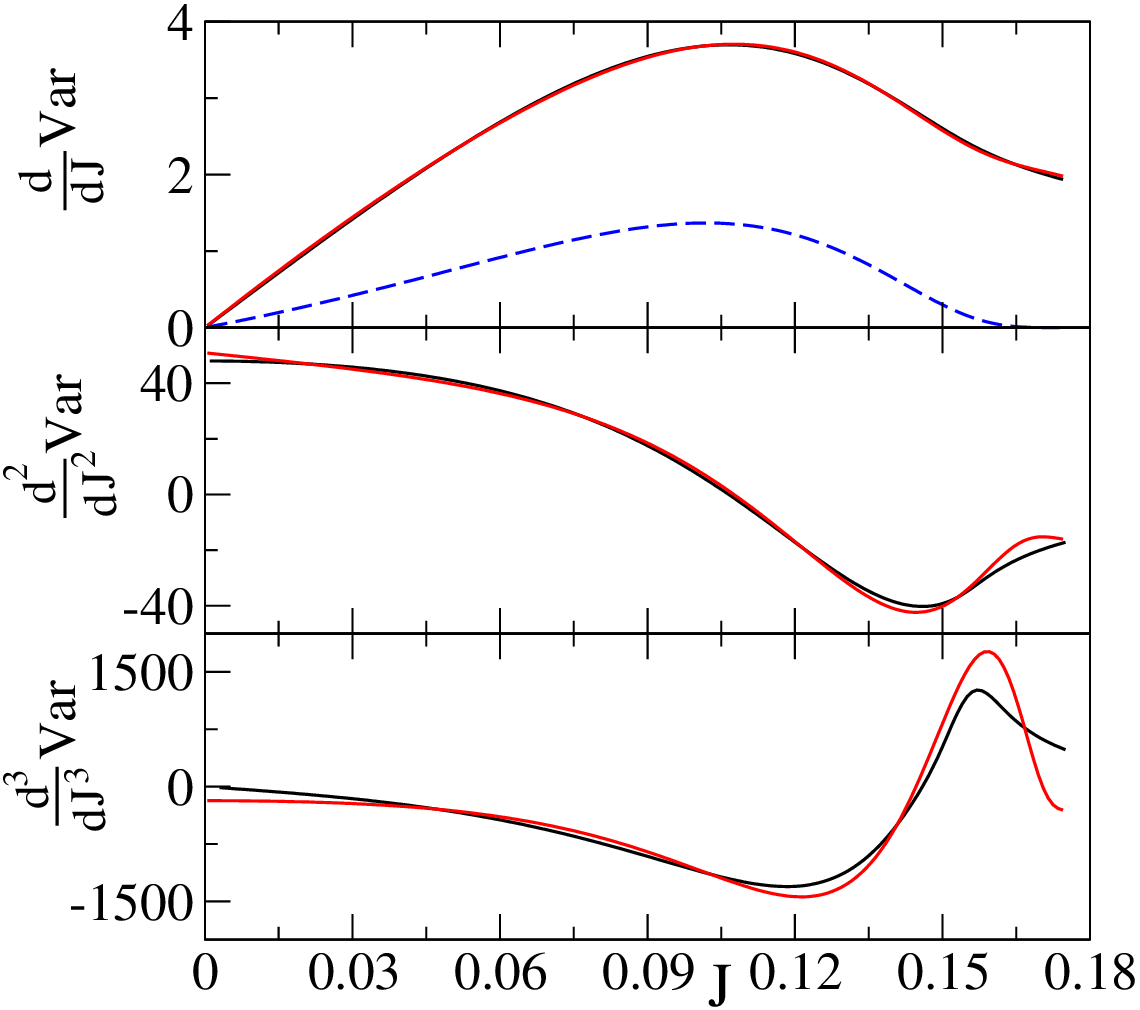}
\caption{
Comparison between 
numerics  and the BKT fit for 
 the double filling factor ($n=2$).
Black lines show numerics   for the
$M=800$ system. 
Red lines follow from 
 (\ref{dVar_ABCD}) 
evaluated with the coefficients from (\ref{ABCD_n2}).
 The dashed blue line in the top panel depicts the universal 
part of the fitted function.
}
\label{kt_n=2}
\end{figure}

The results for the filling factor $n=2$ are presented 
in Figs. \ref{d1n2}--\ref{kt_n=2}.
A quick comparison of 
Figs. \ref{d1n1}--\ref{kt_n=1} 
to Figs. \ref{d1n2}--\ref{kt_n=2}
shows that  qualitative features of 
derivatives of the variance
are the same for filling factors $n=1,2$.
As a result of that, we will 
just briefly  summarize below  
quantitative features of 
the double filling 
factor results.

To begin, the relevant  
 Rayleigh-Schr\"odinger perturbative expansion in the
tunneling coupling
now reads \cite{BDNJP2015}
\be
\begin{aligned}
\Var(J)&= 24 J^{2} -192 J^{4} -\frac{396832}{63} J^{6} \\
&+\frac{6770645594}{496125} J^{8} 
-\frac{32931564509156}{9359398125} J^{10}\\
&+\frac{7350064303936751836656911}{173664334164234375} J^{12}
 + O\B{J^{14}}.
\end{aligned}
\label{var_n=2}
\ee

Extrema of numerical results, 
for the $M=800$ system,
 are characterized by
\begin{equation}
\begin{array}{c|c|c}
  & \text{minimum} & \text{maximum}   \\
  \hline
  d\Var/dJ      &      &   0.107       \\
  \hline
  d^2\Var/dJ^2  & 0.146  &        \\
  \hline
  d^3\Var/dJ^3  & 0.118   &  0.157
\end{array}
\end{equation}
while the ones following from   expansion (\ref{var_n=2}) are
\begin{equation}
\begin{array}{c|c|c}
  & \text{minimum} & \text{maximum}   \\
  \hline
  d\Var/dJ      &       &   0.106    \\
  \hline
  d^2\Var/dJ^2  &  0.157  &        \\
  \hline
  d^3\Var/dJ^3  &  0.128   &
\end{array}\,.
\label{jjeeaa}%
\end{equation}
The position of the maximum of the third derivative is not listed
in (\ref{jjeeaa})
 because such a maximum is absent in
 the third derivative of (\ref{var_n=2}).

Fitting of (\ref{dVar_ABCD}),
to  numerical data for the $M=800$ system 
in the range $0\le J\le0.175$,
has been  done with $J_c=0.18$.
Such a value of $J_c$ has been
taken from 
the survey presented in
Ref. \cite{KrutitskyPhysRep2015}.
We have obtained 
\be
\begin{array}{c|c|c|c}
 A & B & C & D \\
 \hline
-13.2(2)  & 1.134(2)   & 38.7(3)  & -157(2) 
\end{array}\,.
\label{ABCD_n2}
\ee
This time, however, the 
result for the parameter $B$ cannot be 
compared to the previous studies because we are 
unaware of any reference reporting it.

\section{Numerics}
\label{Numerics_sec}

The ground state calculations have been performed using 
implementation
of the 
Density Matrix Renormalization Group (DMRG) algorithm \cite{Schollwock2011}
provided by the 
iTensor  package \cite{Fishman2020}. 
The numerical
method minimizes the mean energy of  variational 
many-body ground states
expressed in the Matrix Product State (MPS) form. That
representation is given by 
\be
|\psi_{\textrm{MPS}}\rangle=
\sum_{i_1,i_2,\ldots,i_M} 
A^{[1]}_{i_1}
A^{[2]}_{i_2}\cdots A^{[M]}_{i_M} |i_1,i_2,\ldots,i_M\rangle,
\ee
where  $A^{[m]}_{i_m}$ are 
$1\times \chi$, $\chi\times\chi$, and $\chi \times 1$
matrices for $m=1$, $1<m<M$, and $m=M$,  respectively. 
The index $i_m=0,1,\dots,7$ represents the
on-site population of  the $m$-th lattice site (we have checked that 
such a choice leads to well-converged results).
The set of all states, for the given 
Schmidt dimension
$\chi$, 
forms a variational manifold.

The MPS representation is exact  for  large-enough
$\chi$.  We have used 
 $\chi=400$ for $M\le400$ and 
$\chi$ up to $1600$ for  $M=800$. 
This concerns simulations at both the unit and double filling factors.
Too small $\chi$ results in bad convergence
of ground states, which translates into noise complicating calculations
of derivatives.
The sufficiently-large  $\chi$ grows with
$M$, making investigations of larger systems 
prohibitively expensive in terms
of time and computer resources. 

We have  monitored
the quality of our simulations by the study of 
 discarded weights $w_m=\sum_{j>\chi} (\lambda_j^{[m]})^2$,
 where $\lambda_j^{[m]}$ are Schmidt coefficients \cite{Schollwock2011,Fishman2020}. 
 All simulated states have been converged down to
$w_m\leq10^{-10}$ for all $1\le m\le M$. 
The iTensor
``cutoff'' parameter, used for DMRG internal linear algebra truncation,
has been  set to $10^{-13}$ \cite{Fishman2020}.

The variance has been  computed from 
ground states, generated  by the above-mentioned procedure
for  $J=\{J_i\}$, where $J_{i+1}-J_i=\delta\ll1$.
In order to minimize 
influence of open boundary conditions on our results,
we have evaluated it at the central lattice site:
$\Var=\la\hat n^2_{M/2}\ra -n^2$, where $n=1,2$.
Its numerical derivatives have been 
obtained  from  the symmetric prescription
\begin{align}
&\frac{d}{dJ}\Var(J_{i+1/2})\approx\BB{\Var(J_{i+1})-\Var(J_{i})}/\delta,\\
&\frac{d^2}{dJ^2}\Var(J_{i})\approx\BB{\frac{d}{dJ}\Var(J_{i+1/2})
-\frac{d}{dJ}\Var(J_{i-1/2})}/\delta,
\end{align}
etc., where $J_{i+1/2}=J_i+\delta/2$.

From these formulae, one easily sees that the key limitation of such a procedure 
follows from the fact that the  denominator
of the $n$-th order derivative is given by $\delta^n$. This implies that 
reliable results are obtained only when 
accuracy of determination of 
$\Var(J_{i})$ is much better  than $\delta^n$. To compute the first and second
derivative of the variance, we have used $\delta=0.001$ getting  smooth results. 
However, our results for the third derivative, obtained with such $\delta$, exhibit   
small  fluctuations near the critical point 
due to worse
accuracy of determination of the variance there.
The problem with 
smoothness of the  third derivative has been  resolved by 
employment of $\delta=0.002$, which does not harm the overall
accuracy of our studies as such $\delta$ is still sufficiently small.

Alternatively, one could have solved such an issue by 
differentiation of  a smooth curve that has been 
fitted to $\Var(J_i)$ data.
We have not explored this option 
 here because the above-mentioned procedure 
 straightforwardly delivers 
good-quality  results.

Finally, at the risk of stating the obvious,
we mention that whole discussion from this appendix 
applies to our studies of both the unit and double filling factor 
systems.


\begin{thebibliography}{53}
\bibitem[{rem({\natexlab{b}})}]{remarkNext1}
\bibinfo{note}{Note that the non-zero value of derivatives of the variance at
  the critical point, suggested by extrapolation of the results from Figs.
  \ref{d1n1}--\ref{d3n1} to $J=J_c$, cannot be attributed to finite-size
  effects, which are too small. The same can be said about the double filling
  factor case, which is depicted in Figs. \ref{d1n2}--\ref{d3n2}.}
\end{thebibliography}

\end{document}